\documentclass[twocolumn, amsmath, superscriptaddress, amsfonts,prb]{revtex4-2}

\usepackage{graphicx}
\usepackage{lipsum}

\pdfoutput=1

\begin{document}

\title{Threshold switching in solar cells and a no-scribe photovoltaic technology }

\author{Victor G. Karpov}\email{victor.karpov@utoledo.edu; Author to whom correspondence should be addressed.}\affiliation{Department of Physics and Astronomy, University of Toledo, Toledo,OH 43606, USA}
\author{Diana Shvydka}\email{diana.shvydka@utoledo.edu}\affiliation{Department of Radiation Oncology, University of Toledo Health Science Campus, Toledo, OH 43614, USA}
\author{Sandip S. Bista}\email{sandip.bista@rockets.utoledo.edu}\affiliation{Department of Physics and Astronomy, University of Toledo, Toledo,OH 43606, USA}

\begin{abstract}
We show that thin film CdTe solar cells exhibit the phenomenon of threshold switching similar to that in phase change and resistive memory. It creates a conductive filament (shunt) through the solar cell reaching the buried electrode, such as the transparent conductive oxide (TCO) in CdTe based photovoltaics (PV). While in the existing PV the buried electrode was routinely contacted via laser scribe filled metals, our work paves a way to an alternative technology of no-scribe PV.
\end{abstract}

\maketitle

\begin{figure}[b!]
\includegraphics[width=0.45\textwidth]{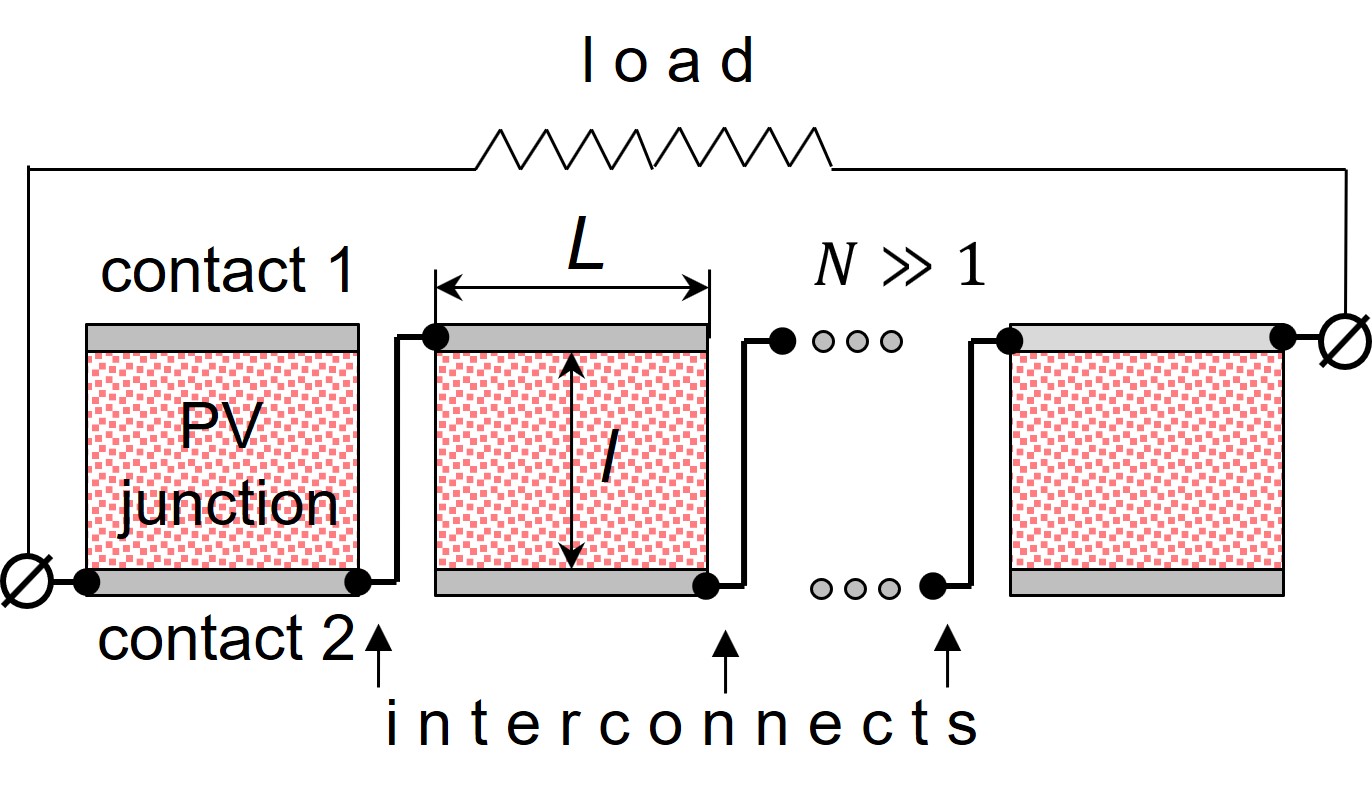}
\caption{A conceptual design of an integrated PV module with $N\gg 1$ individual cells of width $L$ each assembled in series by means of the interconnects between the opposing electrodes of neighboring cells.  The interconnects are located in narrow grooves created by laser scribing.
\label{Fig:Module}}
\end{figure}

A common feature of various large area PV is the presence of interconnects organizing individual solar cells into integrated modules as sketched in Fig. \ref{Fig:Module}. Connections between the opposite electrodes of neighboring cells arrange them into series utilizing the integral voltage. Such interconnects require electric contacts with a buried (bottom) electrode and are created in voids produced by laser scribing \cite{compaan2000, scheer2011,annigoni2019,booth2010,taheri2021,giacomo2020,moon2015,schneider2006,gecysl2017,stelmaszczyk2014} in the initially deposited continuous multi-layer structures. In addition to the cost concerns, laser scribed interconnects raise reliability issues. 

The cell dimension $L$ is determined by the compromise between the cell resistive voltage loss $\delta V$ (increasing with $L$) and the current loss (increasing with cell number $N$ and thus decreasing with $L$). More quantitatively, \cite{karpov2001} \begin{equation}\label{eq:celllength}L=\sqrt{\delta V/sj}\end{equation} where $s$ and $j$ are respectively the electrode sheet resistance and the current density (A/cm$^2$). For 2D roundish cells, Eq. (\ref{eq:celllength}) states that the current $jL^2$ produces resistive voltage drop $\delta V=jL^2s$.  Eq. (\ref{eq:celllength}) holds as well for 1D cells with $s$ and $j$ understood as the resistance and current per length. Using the typical $j\sim 20$ mA/cm$^2$, $s \sim 10$ Ohm, and $\delta V\sim 0.03$ V, $L$ cannot exceed one centimeter, and each module must have many scribes.

\begin{figure}[b!]
\includegraphics[width=0.3\textwidth]{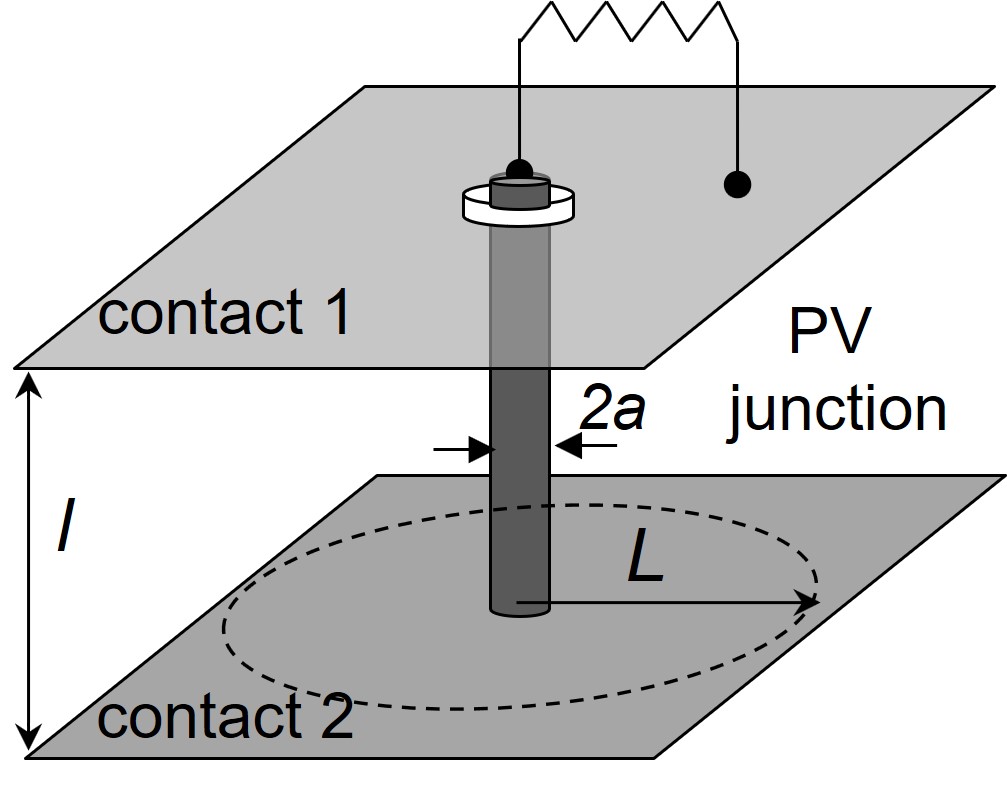}
\caption{The conceptual point contact design without scribes and interconnects. The electric current from proximity of radius $L$ is collected by connecting the bottom electrode through  a low resistive conducting filament insulated from the top electrode (white ring in the diagram); $l$ is the device thickness.
\label{Fig:Noscribe}}
\end{figure}

Here, we propose a physical concept allowing electric contact with the buried electrode without scribing. The visual representation of it in Fig. \ref{Fig:Noscribe} appears rather trivial: a conducting filament (`shunt' in PV jargon) to the bottom electrode with its tip used as the terminal. It collects current from a circular area of radius $L$. A set of such filaments separated by distances of the order of $L$ will then collect current from the entire PV module.

In our work, the conductive filaments are created by threshold switching, which is known to be a well controlled process with a defining property of a drastic decrease in resistance under electric fields exceeding a certain threshold value. Along with the chalcogenide films, \cite{ovshinsky1968,adler1978,petersen1976,karpov2008,kribs2009,raoux2010} there are many other materials, in particular various metal oxides, exhibiting the  phenomenon of threshold switching.\cite{rram,zahoor2020,wang2020,yu2014}  The characteristic switching fields vary between these materials in the ballpark of $\sim 10^4-10^6$ V/cm, which, for device thicknesses $l\sim 1-10$ $\mu$m translates into convenient switching voltages of $\sim 1-100$ V.

The above mentioned materials are not photovoltaic. Threshold switching has been observed as well with some materials related to PV: amorphous and polycrystalline silicon, \cite{owen1983,mahan1982} CIGS, \cite{dongaonkar2012}, CdZnTe, \cite{zha2015} and perovskites. \cite{ignatiev2006} Our work here demonstrates the phenomenon of threshold switching in a functional PV structure rather than in a particular material component.

The microscopic mechanisms of threshold switching vary between different systems. They typically evolve through the formation of a narrow conducting filament via either crystallization or rearrangements of certain structural defects, impurities, or electric dipoles. \cite{ovshinsky1968,adler1978,petersen1976,karpov2008,kribs2009,raoux2010,rram,zahoor2020,wang2020,yu2014,karpov2007,karpov2017} The filament can be thermodynamically stable or long-lived metastable. \cite{karpov2011} It appears tunable with the reciprocal area $1/\pi a^2$ and resistance $R$ proportional to the maximum (compliance) current $I_0$ flowing in the course of switching. \cite{adler1978,petersen1976,karpov2017,karpov2011}

\begin{figure}[hbt]
\includegraphics[width=0.48\textwidth]{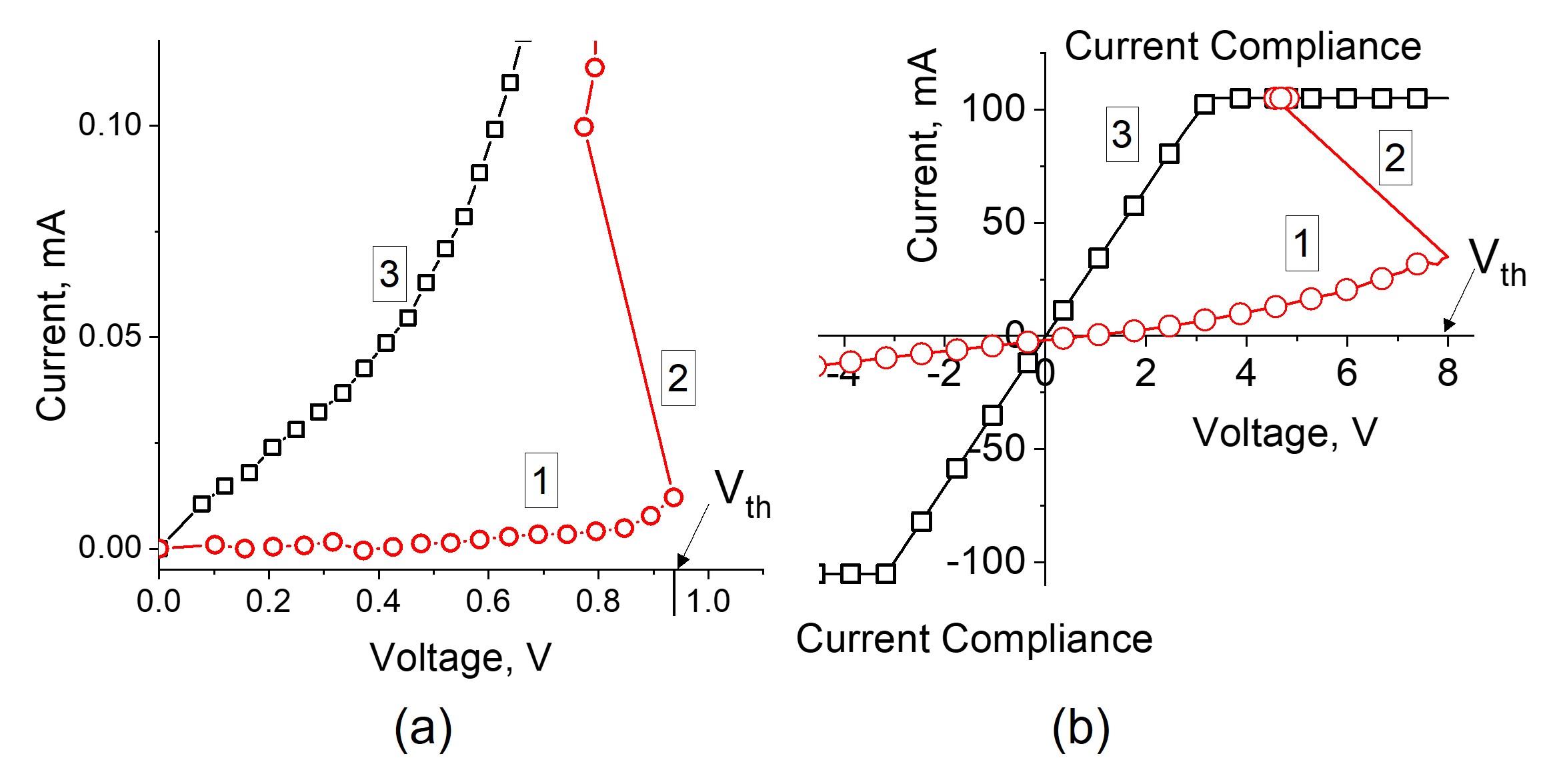}
\caption{(a) Switching (breakdown) event in a phase change memory chalcogenide material (GST) revealed in the I-V snap-back feature. The marked branches correspond to high resistance phase (1), snap-back region (2), and low resistance filament (3). The data are from Ref. \cite{karpov2007}. (b) Similar I-V including snap-back related to the formation of a low resistance filament in a CdTe based solar cell. The branch marking is the same as in (a).
\label{Fig:PCMPV}}
\end{figure}

We recognize the phenomenon of threshold switching based on its distinctive snap-back feature in the current-voltage characteristics presented in Fig. \ref{Fig:PCMPV} and reflecting the drastic resistance drop. The snap-back of Fig. \ref{Fig:PCMPV}(a) is typical of all known threshold switching systems. We observed a similar snap-back evidence in a solar cell as presented in Fig. \ref{Fig:PCMPV} (b). The measured threshold voltage $V_{th}\sim 10$ V is qualitatively consistent with $V_{th}\sim 1$ V for the phase change memory possessing the film thickness that is by an order of magnitude smaller.

We observed that subsequent voltage sweeps in the switched solar cells produced I-V curves fully overlapping with the branch 3 in Fig. \ref{Fig:PCMPV} (b). An obvious interpretation is that, since formed, the conductive filament remains intact forming a stable contact with the buried electrode. Furthermore, the slope of branch 3 in Fig. \ref{Fig:PCMPV} (b) is consistent with the value of the second term in Eq. (\ref{eq:ohmres}) below assuming $L/a\sim 1000$, which points at $a\sim 1$ $\mu$m, in the ballpark of  values directly measured for comparable film thickness. \cite{adler1978,petersen1976} Therefore, the switching produced filaments in solar cells are truly low resistive and capable of serving as good contacts to TCO.

A theoretical description of the filament current collection is rather short extending from Eq. (\ref{eq:celllength}) to Eq. (\ref{eq:ohmres}) below.
Modifying the known 1D result \cite{karpov2001} to 2D case, it is straightforward to show that the screening length $L_R$ for a finite shunt resistance $R$, is given by,
\begin{equation}\label{eq:finres}L_R=L\sqrt{s/(R+s)}\end{equation}
Therefore, the area of current collection shrinks with $R$.

The total current $I$ through the filament corresponds to the electrode surface current density $j=I/2\pi r$ at distance $r$ from the filament. The corresponding electric field  $E=sj=Is/2\pi r$ yields the potential difference, $\int ^L_aEdr=(Is/2\pi)\ln(L_R/a)$
where we have approximated the upper limit of integration with the distance $L_R$ within which the current is collected. Adding here the electric potential $IR$ across the filament, the proportionality coefficient between the latter sum and the current yields the total ohmic resistance,
\begin{equation}\label{eq:ohmres}R_{\rm tot}=R+(s/2\pi)\ln (L/a).\end{equation}
Because $a\propto\sqrt{I_0}$, fitting the dependencies $R_{\rm tot}(I_0)$ with Eq. (\ref{eq:ohmres}) can determine the filament parameters.

\begin{figure}[b!]
\includegraphics[width=0.5\textwidth]{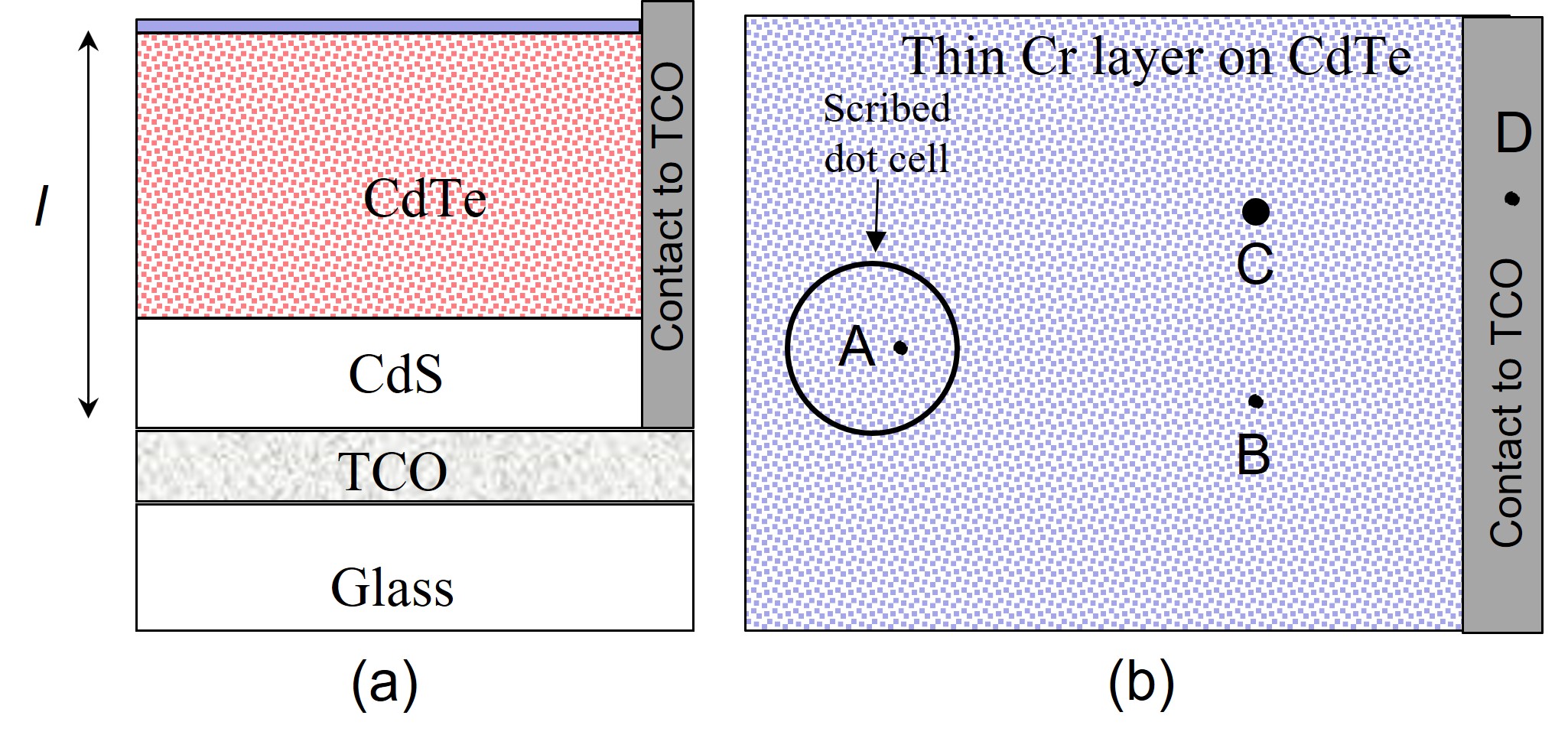}
\caption{(a) Cross-sectional structure of thin film CdTe based PV (not to scale). (b) Top view of the same structure with  points A-D designated for electric contacts: on a back metal (A), on the high resistive electrode (B), on the TCO contact (D); point C belongs in the high resistive electrode and is undergone switching that creates a low resistance path from that point to TCO. (Note that the physical connection between Cr and extrinsic contact to TCO is electrically insignificant due to high sheet resistance of Cr.)
\label{Fig:CdTenoscribe}}
\end{figure}

Our experimental verification is aimed at showing how the switching created filament can be used to contact the buried electrode and thus measure the current-voltage (I-V) characteristics of a solar cell. Depicted in Fig. \ref{Fig:CdTenoscribe} (a) and (b) are correspondingly the cross-sectional structure and a top view of our used CdTe based polycrystalline solar cells exhibiting weak p-type without any significant doping. Our cells were produced by vapor transport deposition; the details given in Refs.  \cite{shvydka2002,shvydka2002a}. Table \ref{tab:exp} and Fig. \ref{Fig:CdTenoscribe} (b) list the contact points utilized in our experiments and their expected results.

\begin{figure}[thb]
\includegraphics[width=0.48\textwidth]{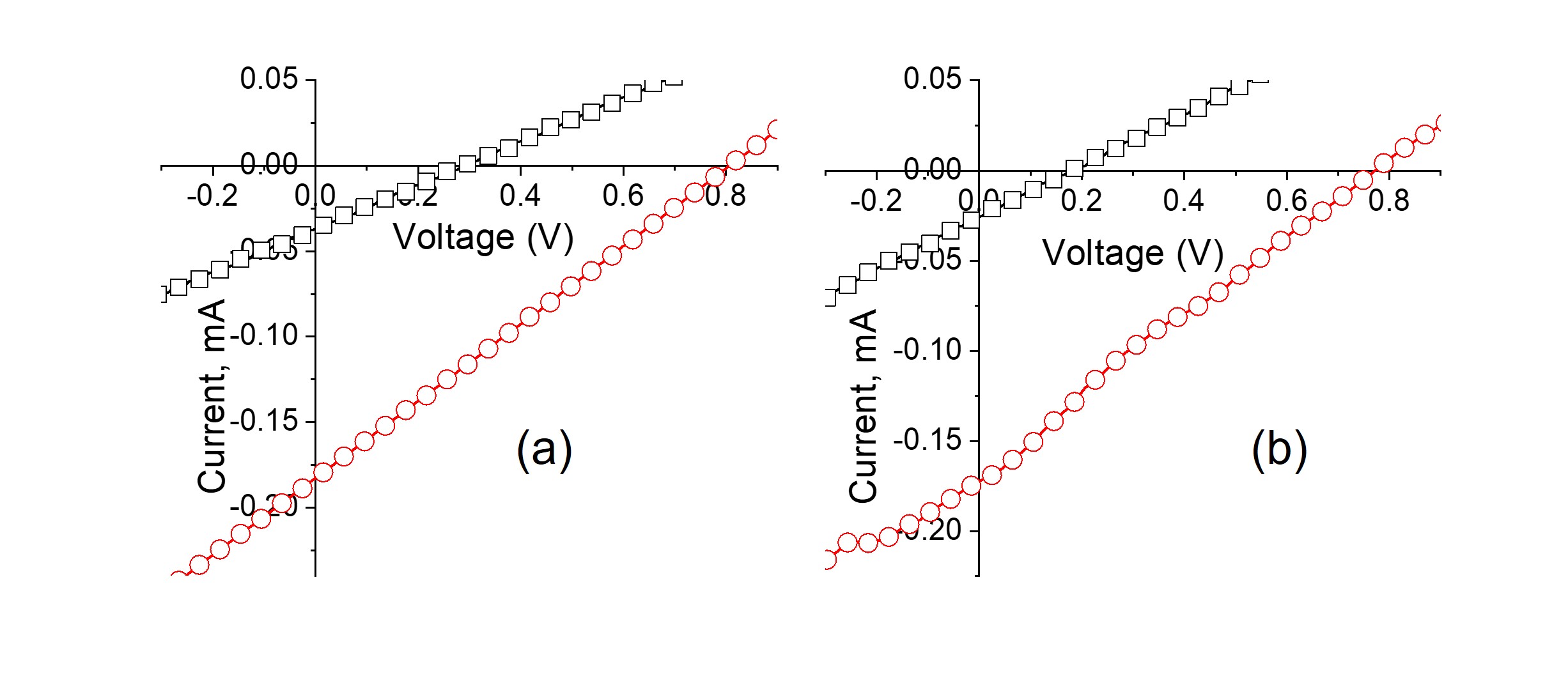}
\caption{(a) Light and dark IV characteristics measured between an arbitrary contact point B on a continuous thin Cr contact and D. (b) Light and dark IV characteristics measured between the contact points C and D as described in Table \ref{tab:exp} where location C had undergone switching.
\label{Fig:TCOshunt}}
\end{figure}

\begin{figure}[bht]
\includegraphics[width=0.48\textwidth]{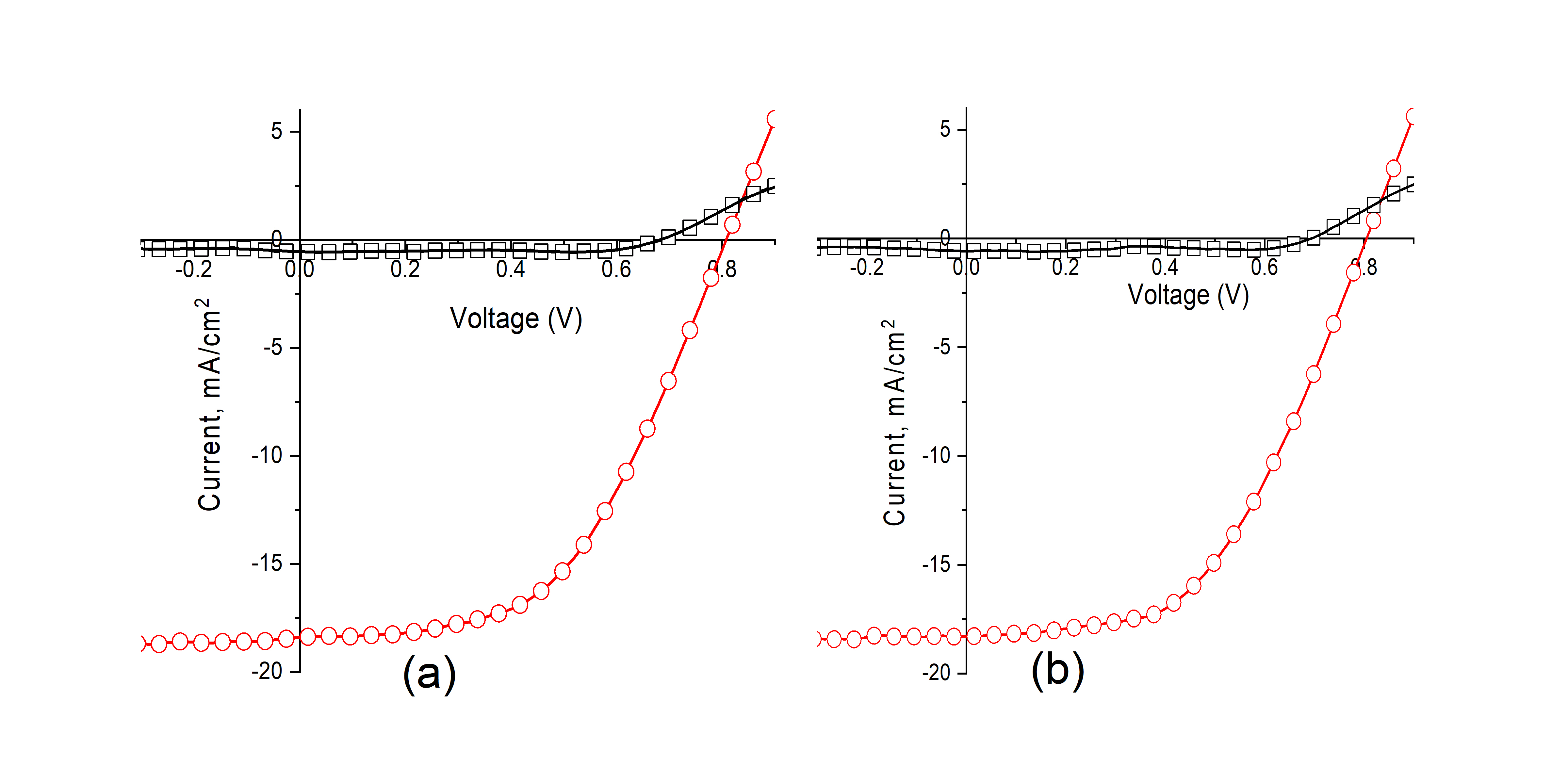}
\caption{(a) Light and dark characteristics of a small cell scribed through the thin Cr layer measured between the locations A and D. (b) Light and dark characteristics of a small cell scribed through the thin Cr layer measured between the locations A and C after the location C had undergone switching.
\label{Fig:CellShunt}}
\end{figure}

Unlike of the standard CdTe cells, our design contains a 5 nm sputtered thin Cr layer of high sheet resistance of $s\approx 1.2$ k$\Omega$, \cite{shvydka2003} almost 100 times higher than the standard TCO sheet resistance used with our cells, $s=15$ $\Omega$. Therefore the characteristic lengths in Eq. (\ref{eq:celllength}), (\ref{eq:finres}) become by an order of magnitude shorter than for cells without the Cr layer. The cells with thin Cr layer are similar to the standard
ones, except that TCO now plays the role of the low resistance electrode. \cite{shvydka2003} Note that because of the high sheet resistance, there is a significant ohmic loss, so that the current in large cells becomes close to ohmic: $I\approx (V-V_{oc})/s$ where $V_{oc}$ is the open circuit voltage as was observed (cf. Fig. \ref{Fig:TCOshunt}). For small cells, in the millimeter range, scribed through the Cr film, the series sheet resistance is not so important, and I-V characteristics demonstrate more standard shapes (Fig. \ref{Fig:CellShunt}).

The role of high-resistive Cr electrode here is that it allows one to electrically contact the CdTe surface without much addition to the contact series resistance (unlike the contacts of many standard electrodes with bare CdTe surface known to be electrically blocking and/or unstable, such as Ni, Al, Cu). Another useful feature of the high resistive electrode is that it prevents significant lateral spreading and parasitic leakage of electric currents. While it has been established earlier that Cr tends to form a Schottky barrier on CdTe surface, \cite{dharmadasa1988}, the most significant factor behind using Cr here was that it formed extremely strong adhesion layer on CdTe surface providing excellent electric connection, unique among other metals tried.

The above declared property of conductive filaments forming good contacts with TCO was verified by comparing the I-V characteristics read through the switched location [C in Fig. \ref{Fig:CdTenoscribe} (b)] with those obtained through the standard TCO soldering contact [D in Fig. \ref{Fig:CdTenoscribe} (b)]. Table \ref{tab:exp} presents more in detail specifications of the related measurements presented in Figs. \ref{Fig:TCOshunt} and \ref{Fig:CellShunt}. Conceptually, they prove as well that PV scribes can be replaced by the switching produced conductive filaments.

\begin{table}[hbt]
\caption{Various pairs of points in Fig. \ref{Fig:CdTenoscribe} (b).}
\begin{tabular}{|p{2cm}|p{6cm}|}
   \hline
  Pair of points & Physical meaning of the measurement\\ \hline
 B-D& An arbitrary location B is used to measure the IV curves of a  high resistive electrode solar cell through the standard TCO contact (point D); presented in Fig. \ref{Fig:TCOshunt} (a).  \\
 B-C& Expected  ohmic IV after threshold switching at point C revealing the presence of conductive filament and identical to that read through the standard TCO contact as shown in Fig. \ref{Fig:TCOshunt} (b)\\
 A-D & The standard IV characteristic between points A and D shown in Fig. \ref{Fig:CellShunt} (a) for the comparison with the characteristic between A and C where TCO is contacted through a conductive filament.\\
  A-C& Location C undergone switching and expected IV characteristics should coincide with those of A-D proving the concept as revealed in Fig. \ref{Fig:CellShunt}.\\

   \hline
\end{tabular}
\label{tab:exp}\end{table}

\begin{figure}[th]
\includegraphics[width=0.4\textwidth]{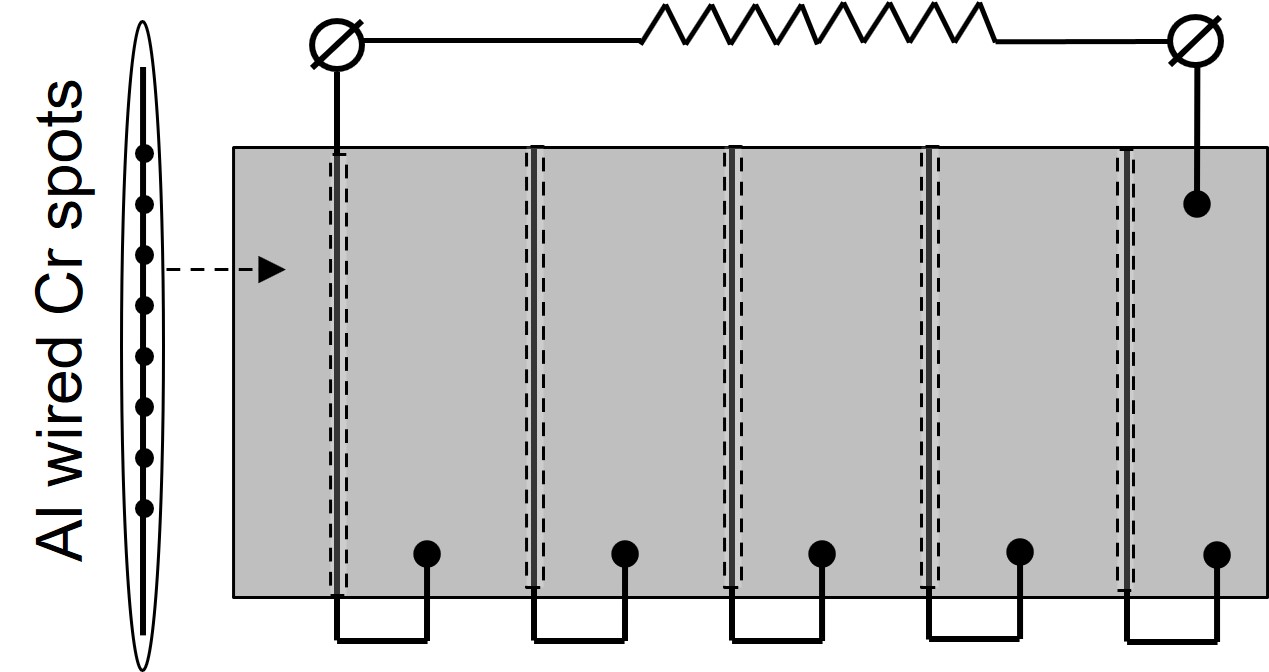}
\caption{A no-scribe module design with vertical lines representing Al wires connecting multiple spots of thin Cr illustrated in the inset. The insulating stripes (dashed rectangles) screen the wires against the subsequent deposition of metal electrode (shown in gray). They  simultaneously make the neighboring cells electrically disconnected. The bold dots represent electric contacts between Al wires and the finishing metal making the  electric circuit equivalent to that of Fig. \ref{Fig:Module}.
\label{Fig:devdes}}
\end{figure}

We would like to briefly mention a few additional observations. (i) In some cases, the threshold switching was so fast that the snap-back I-V branch (2 in Fig. \ref{Fig:PCMPV}) did not show up, yet the switching took place, and the switched device demonstrated straight ohmic branch 3. (ii) We have observed switching under reverse bias, which however was not as reliable showing significant variations from sweep to sweep and a degree of volatility after the bias was removed. (iii) The switching events have been registered as well when the bias was applied within a scribed cell. (iv) Our attempts to observe threshold switching by applying bias to the standard solar cells having certain low resistance back contacts were not successful in general, although in some cases we did observe the reverse bias switching I-Vs with a high degree of variability.

The physics of switching through multi-layer structures of solar cells poses a number of new questions, such as what will be the (time dependent) field distribution through the layers upon bias application, how important might be interference between the capacitive and resistive properties, etc. More work is required to identify the mechanism responsible for the switching phenomena we report in this manuscript.

Finally, implementing the no-scribe technology in  production would require additional developments sensitive to a particular PV brand. For example, keeping an eye on CdTe PV, one can assume the following operations. (1) Depositing and wiring multiple thin-Cr spots on CdTe surface. This is illustrated in the inset of Fig. \ref{Fig:devdes} where Al wiring is assumed as a proven good contact with Cr. (2) Switching material under Cr spots to the conductive state by applying electric bias to the wires. (3) Removing the bias and physically screening the wires against the subsequent back metal deposition. The screening stripes will simultaneously serve as dividers between the neighboring linear solar cells illustrated in Fig. \ref{Fig:devdes}. (4) Depositing back metal and connecting wires to the cells will produce the electric circuit of Fig. \ref{Fig:devdes} equivalent to that of the scribed module.  Various other embodiments can be proposed.

We shall end our discussion by mentioning a possible application of the above observed switching to the important element of various electronics often  termed 'via', \cite{via1,via2} which is an electrical connection between copper layers in a printed circuit board. Typically, a via is a small drilled hole through two or more adjacent layers plated with copper. Our finding here shows that in some cases via can be formed without drilling and plating, by direct application of external bias.

In conclusion, we put forward the concept of threshold switching in solar cells with applications to photo-current collection, and have shown the following.\\
1. Thin film chalcogenide photovoltaics exhibit the phenomenon of threshold switching established earlier for the materials of phase change and resistive memory. \\
2. The switching is most efficient under forward bias resulting in a conductive filament through the device.\\
3. The filaments can serve as contacts with the buried conductive layer (such as TCO in CdTe substrate configuration) in solar cells thus paving a way to the technology of no-scribe photovoltaics.\\

We are grateful to D. Georgiev and V. Borra for useful discussions of our results.\\

The data that support the findings of this study are available from the corresponding author upon reasonable request.

\end{document}